\documentclass[12pt]{article}
\usepackage{graphicx}

\newcommand\pubdate{\today}

\newcommand\pubnumber{LHCb-PROC-2011-041; CERN-LHCb-PROC-2011-041}

\def\PUpsilon      {\ensuremath{\Upsilon}}
\def\Y#1S{\ensuremath{\PUpsilon{(#1S)}}}
\def\OneS  {\Y1S}
\def\TwoS  {\Y2S}
\def\ThreeS{\Y3S}

\def\Title#1{\begin{center} {\Large #1 } \end{center}}
\def\Author#1{\begin{center}{ \sc #1} \end{center}}
\def\Address#1{\begin{center}{ \it #1} \end{center}}

\newcommand\pubblock{\rightline{\begin{tabular}{l} \pubnumber\\
         \pubdate  \end{tabular}}}
\newenvironment{Abstract}{\begin{center}{\bf Abstract}\end{center} \bigskip \begin{quotation}  }{\end{quotation}}
\newenvironment{Presented}{\begin{quotation} \begin{center} 
             PRESENTED AT\end{center}\bigskip 
      \begin{center}\begin{large}}{\end{large}\end{center} \end{quotation}}

\def\beq{\begin{equation}}
\def\eeq#1{\label{#1}\end{equation}}
\def\eeqn{\end{equation}}

\def\beqa{\begin{eqnarray}}
\def\eeqa#1{\label{#1}\end{eqnarray}}
\def\eeqan{\end{eqnarray}}

\let\bar=\overbar

\def\Dslash{\not{\hbox{\kern-4pt $D$}}}
\def\dslash{\not{\hbox{\kern-2pt $\del$}}}

\def\msb{{\bar{\ssstyle M \kern -1pt S}}}

\begin{document}
\begin{titlepage}
\pubblock

\vfill


\Title{$b\rightarrow s \mu^+ \mu^-$ and $b\rightarrow \mu^+\mu^-$ at the LHC}
\vfill
\Author{Giampiero Mancinelli} 
\Address{on behalf of the ATLAS, CMS, and LHCb Collaborations \\ 
\  \\ 
CPPM, Aix-Marseille Universit\'e, CNRS/IN2P3, Marseille, France \\
giampi@cppm.in2p3.fr}
\vfill


\begin{Abstract}
With their 2010-2011 data set, the LHC experiments have started their quest to observe the rare decays $B^0_{s/d}\to \mu^+\mu^-$. This study will provide very sensitive probes of New Physics (NP) effects. NP discovery potential lies as well in the study of the decay $B^0_d \to K^{*0}\mu^+\mu^-$.
Results and perspectives are presented for studies at the LHC of rare B decays involving flavor changing neutral currents. 
\end{Abstract}

\vfill

\begin{Presented}
The Ninth International Conference on\\
Flavor Physics and CP Violation\\
(FPCP 2011)\\
Maale Hachamisha, Israel,  May 23--27, 2011
\end{Presented}
\vfill

\end{titlepage}
\def\thefootnote{\fnsymbol{footnote}}
\setcounter{footnote}{0}
%


\section{Introduction}

A powerful way for the LHC experiments to search for new physics is via non Standard Model (SM) virtual contributions to rare decays of beauty mesons. 
The results reported here for LHCb are based on 37 pb$^{-1}$ of $pp$  collisions at $\sqrt(s)=7$ TeV delivered by the LHC during 2010.

\section{Search for $B^0_{s/d}\to\mu^+\mu^-$}

Within the SM, the dominant contribution to these modes stems from the Z-penguin diagram, 
while the box diagram is suppressed by a factor of $|m_W/m_t|^2$. 
The Higgs annihilation diagram contributes only negligibly (about 1/1000).
These are FCNC modes which, in addition, are also helicity suppressed. Hence the SM expectation is only 3.2 and $0.1 \times 10^{-9}$ for the $B_s$ and $B_d$ mode respectively~\cite{ref:BURAS}. As the error on the theoretical expectations is small, these modes are very attractive as a SM test bench. They are very sensitive to NP with new scalar or pseudoscalar interactions, as well as models with an extended Higgs sector and high $\tan\beta$, as their branching ratio (BR) goes as $(\tan\beta)^6$.
The current best 95\% Confidence Level (C.L.) limits are 43 and $9.1 \times 10^{-9}$ for $B_s$ and $B_d$ to $\mu^+\mu^-$ respectively~\cite{ref:CDF}; these limits are about a factor 10 over the SM expectation.

LHCb is the only LHC collaboration to date which has presented and published a result on the search for these modes, hence its analysis strategy will be described in more detail in the following. $\mu^+\mu^-$ pairs are selected on triggered events with criteria very close to the ones used for the control samples in order to have efficiency corrections as low as possible. The analysis is performed blind. Each candidate is given a likelihood to be signal or background-like in a 2D space formed by the ensemble of geometrical event variables (GL), and by the invariant mass. 
To translate the number of observed candidates into a BR measurement, normalization samples with well measured BRs are used, for which it is necessary to have good knowledge of the relative efficiencies. 
Finally the CLs method~\cite{ref:Read_02} in bins of mass and GL is used to extract the limits.
About 300 events survive the preselection in the mass signal region, defined as $[M_{B_d}-60,M_{B_s}+60]$ MeV/c$^2$. We expect (SM contribution only) about 0.3 $B^0_s\to \mu^+\mu^-$ and 0.04 $B^0_d\to \mu^+\mu^-$ events in 37 pb$^{-1}$ of data. The power of the geometrical discriminant variables is crucial to drastically reduce the rest of the background. Given the excellent muon ID, most of the background is actually due to real muons from semileptonic decays, while the while peaking background from misidentified two-body hadronic B decays is negligible for the integrated luminosity at hand.

\begin{figure}[htp]
\centering
\includegraphics[width=.48\textwidth]{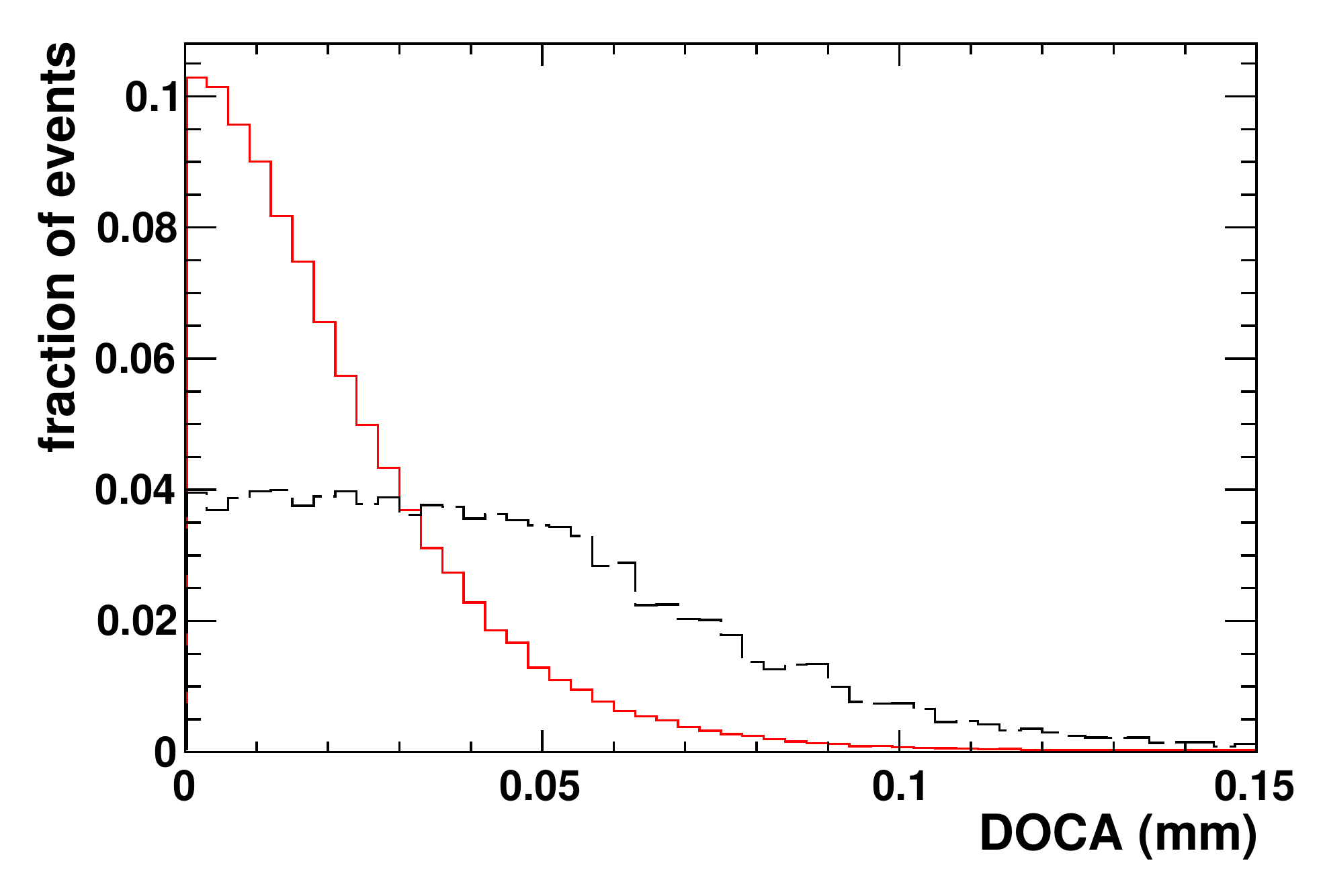}
\includegraphics[width=.48\textwidth]{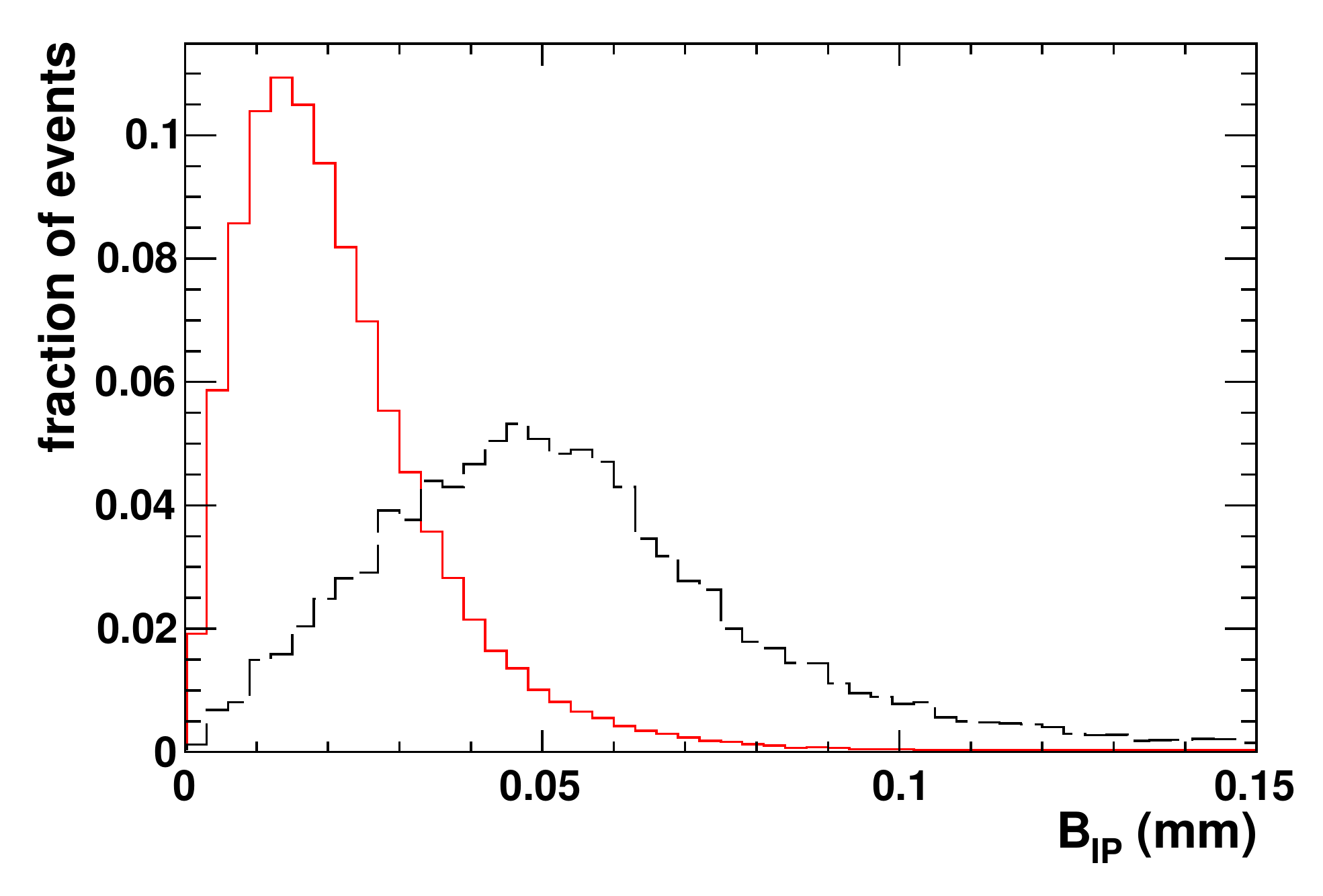}
\includegraphics[width=.48\textwidth]{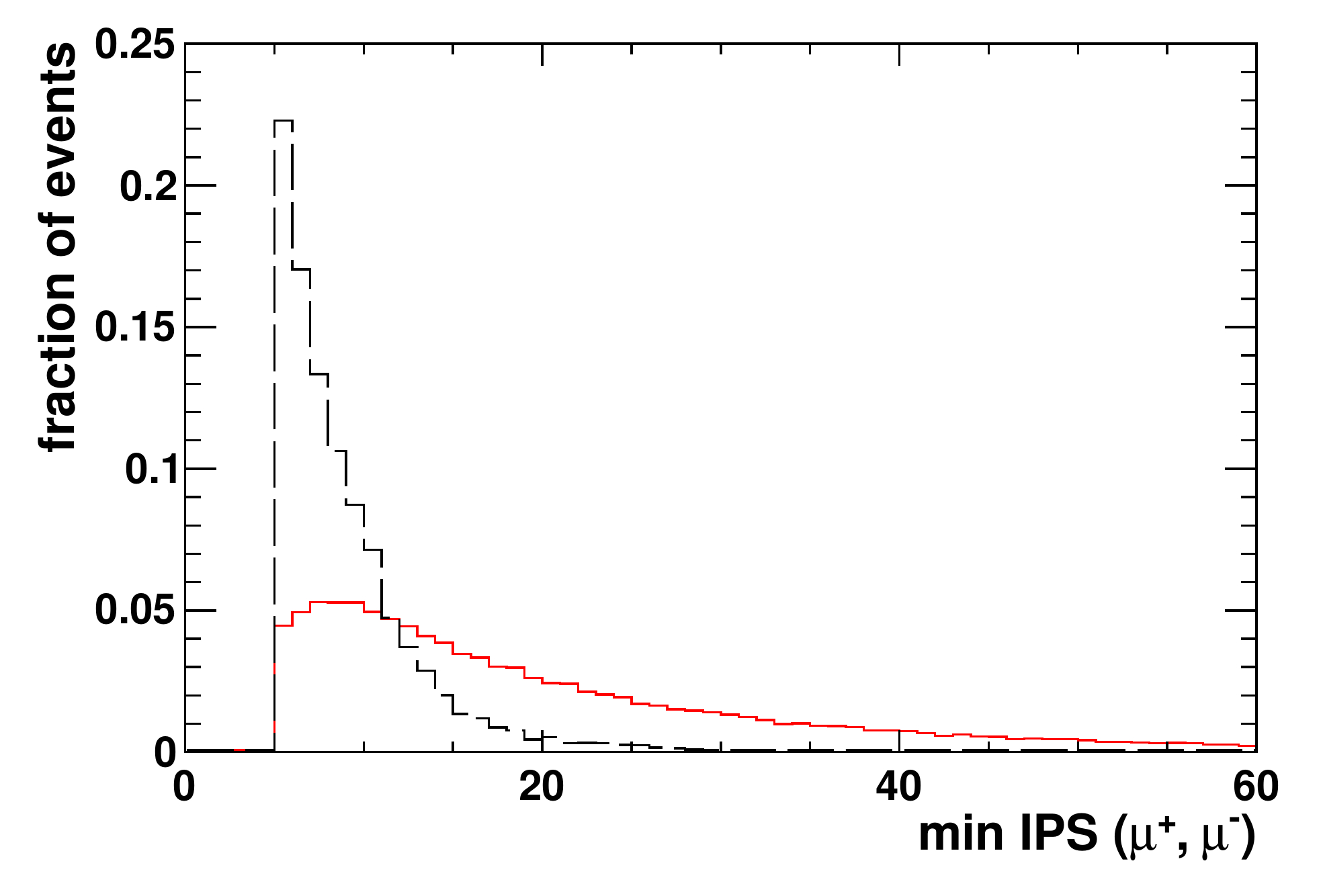}
\includegraphics[width=.48\textwidth]{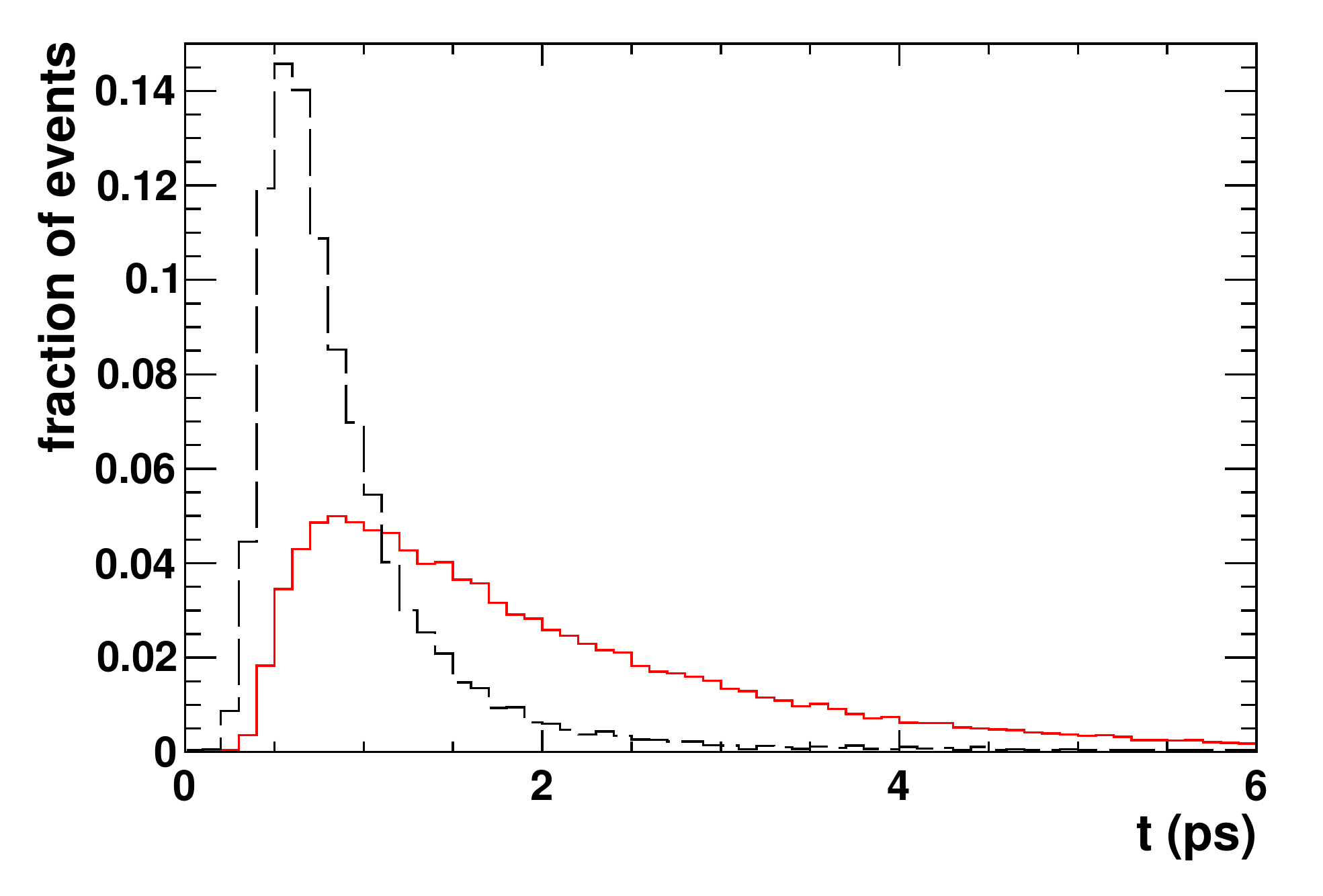}
\includegraphics[width=.48\textwidth]{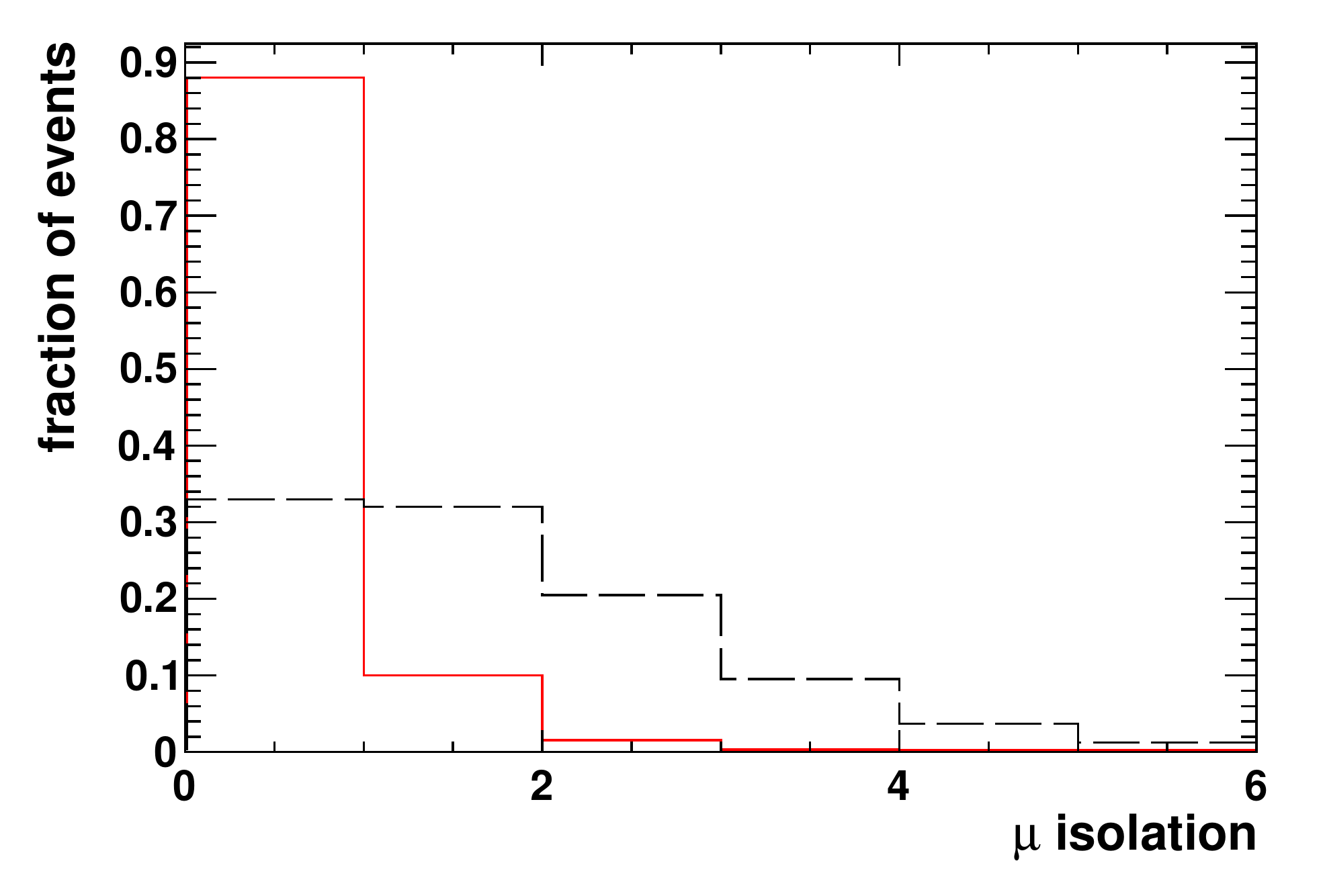}
\includegraphics[width=.48\textwidth]{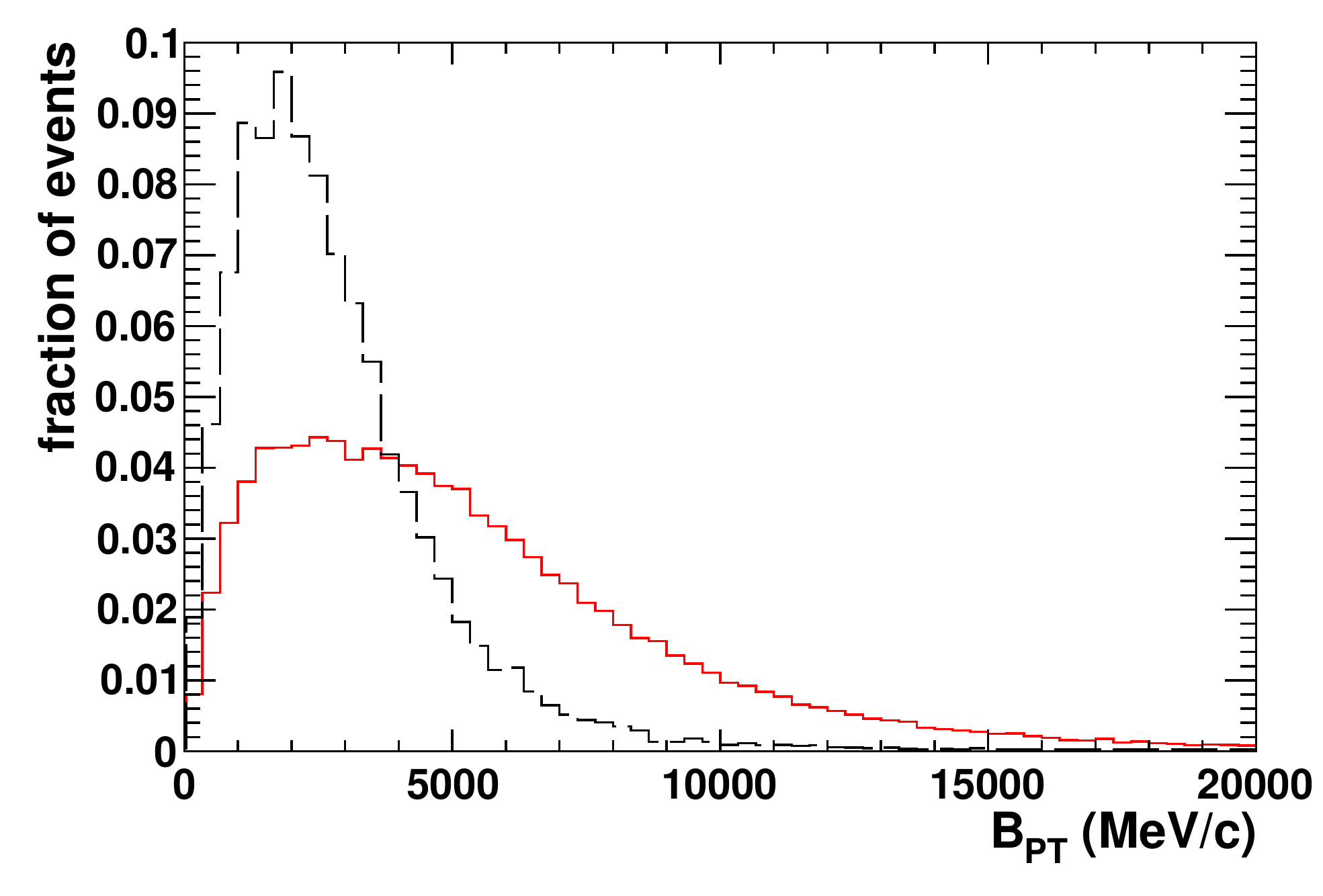}
\caption{Distribution (LHCb), for signal and background, of the variables used in the GL definition. Red: MC $B^0_{s/d}\to\mu^+\mu^-$ signal. Black dashed: MC $b\bar b\to \mu^+\mu^- X$ background.}
\label{fig:gl_vars_1}
\end{figure}

\begin{figure}[htp]
\begin{center}
\includegraphics[width=6cm, height=6cm]{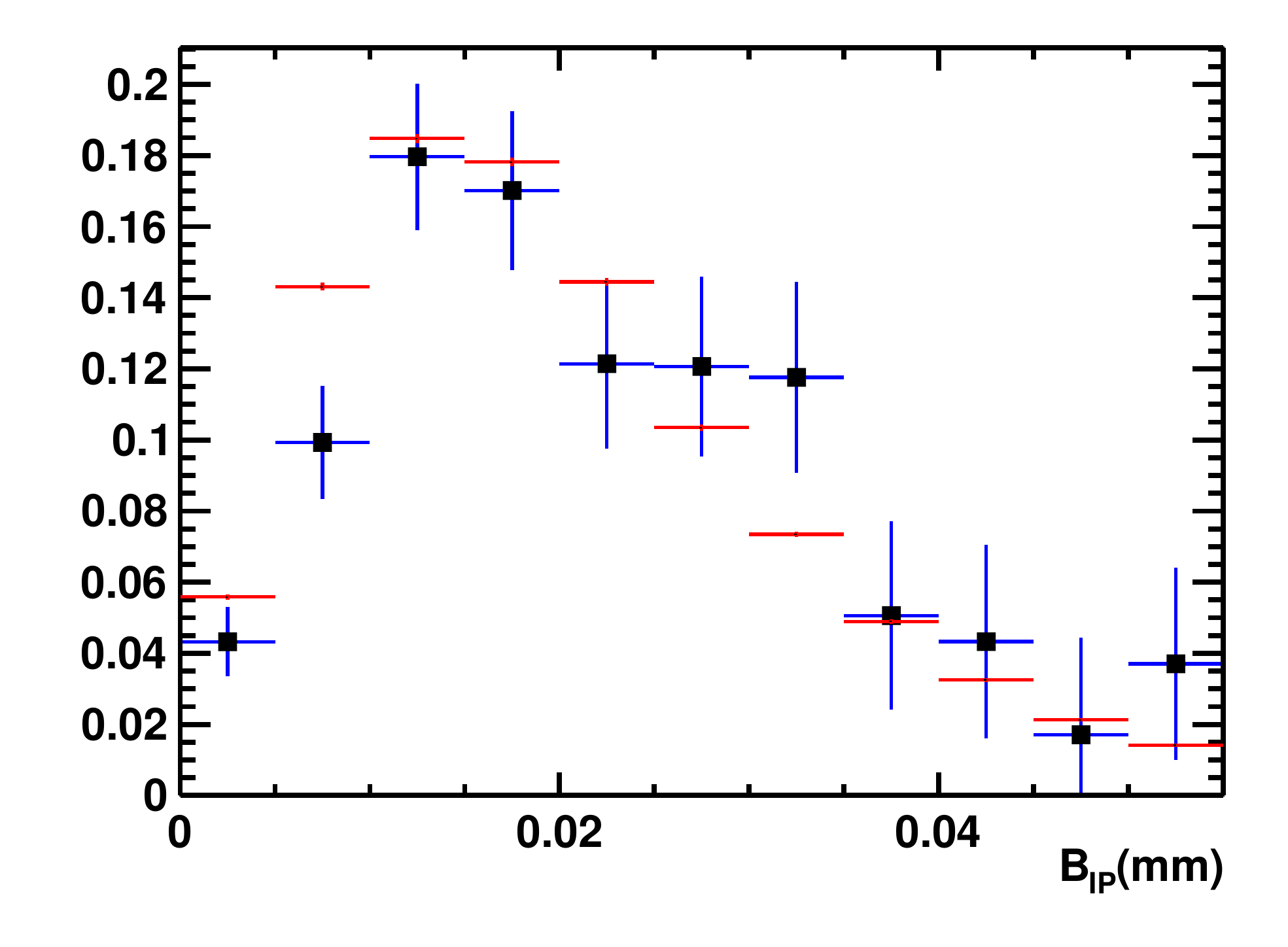}
\includegraphics[width=6cm, height=6cm]{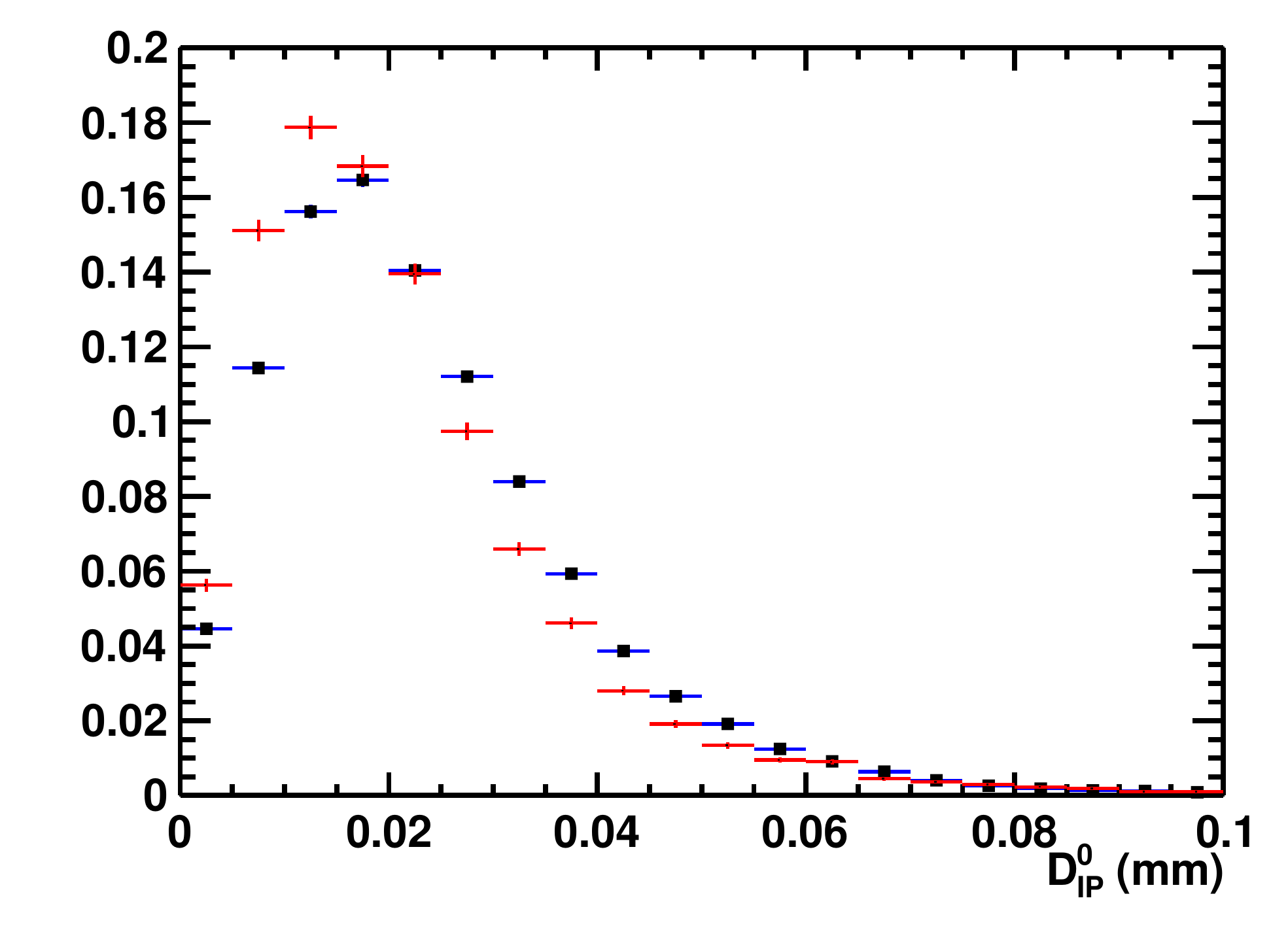}
\includegraphics[width=6cm, height=6cm]{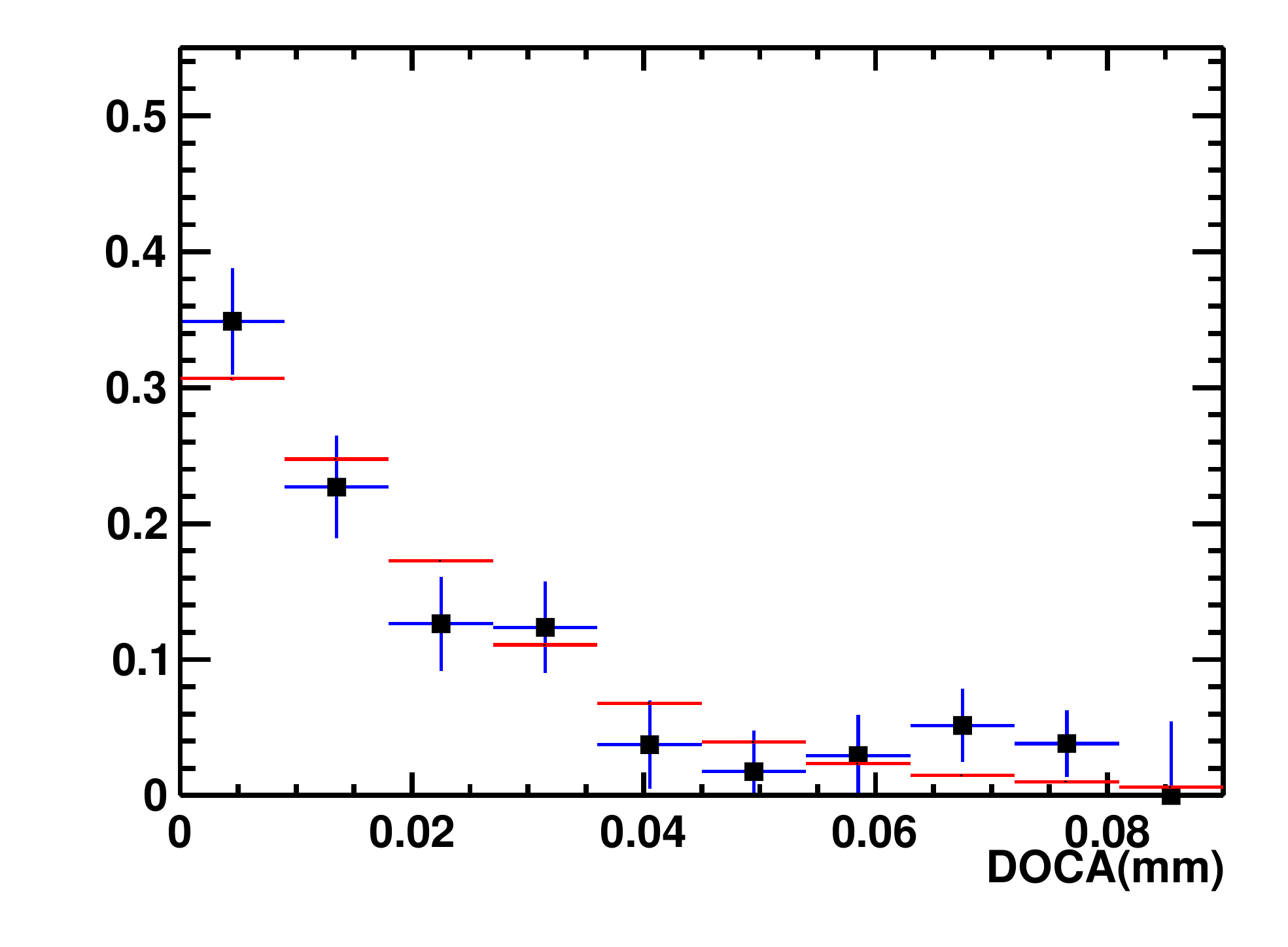}
\includegraphics[width=6cm, height=6cm]{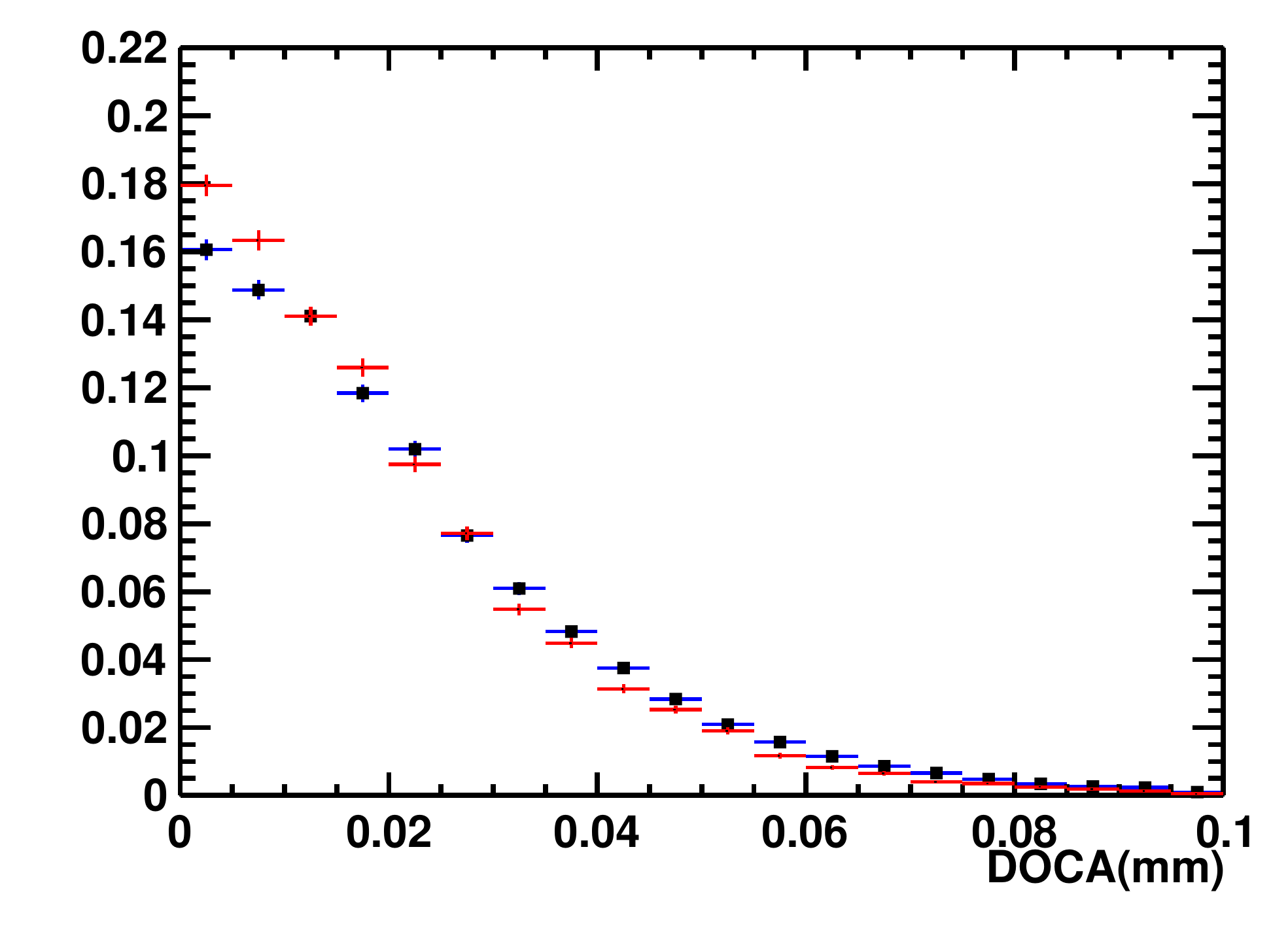}
\includegraphics[width=6cm, height=6cm]{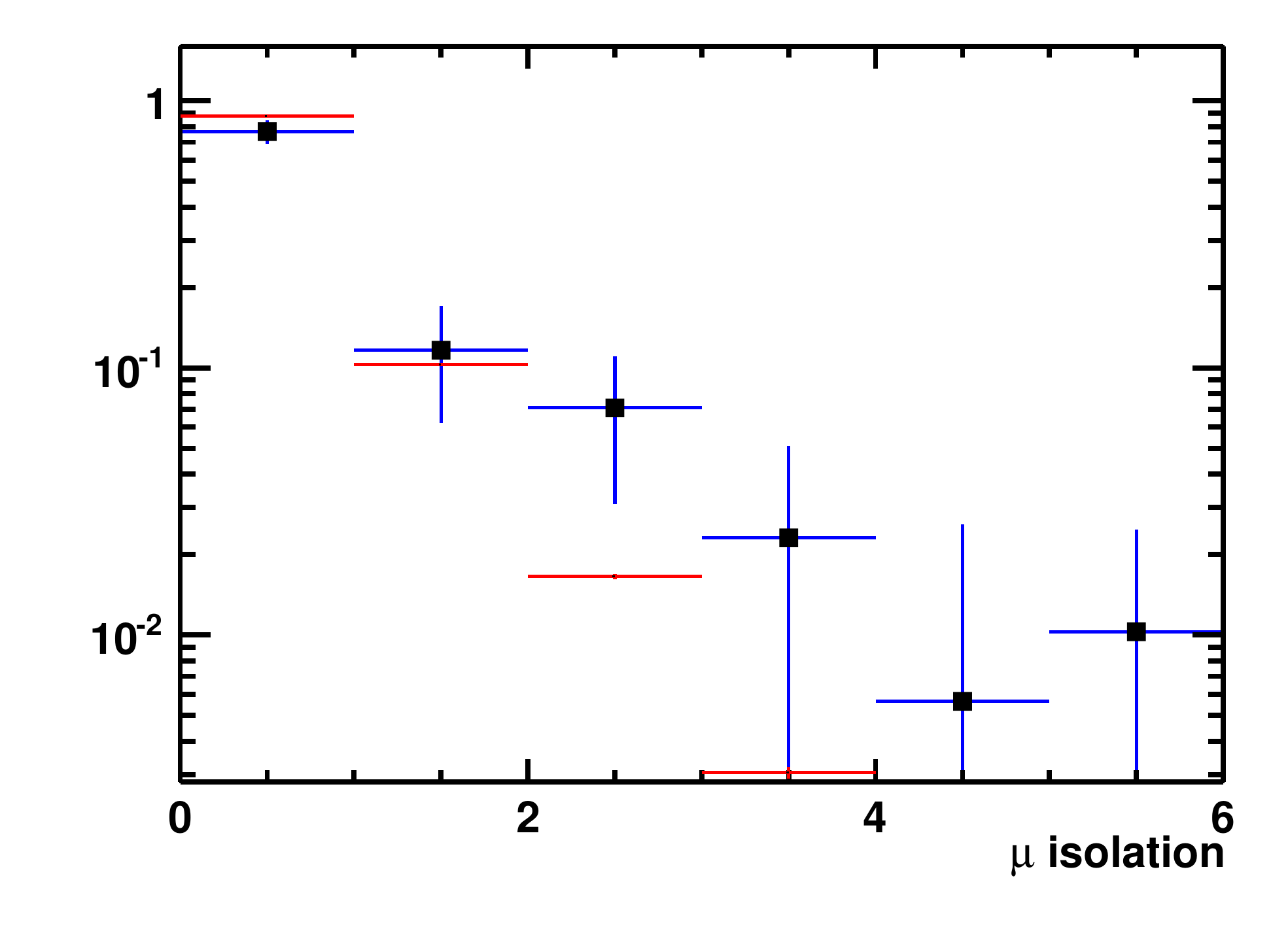}
\includegraphics[width=6cm, height=6cm]{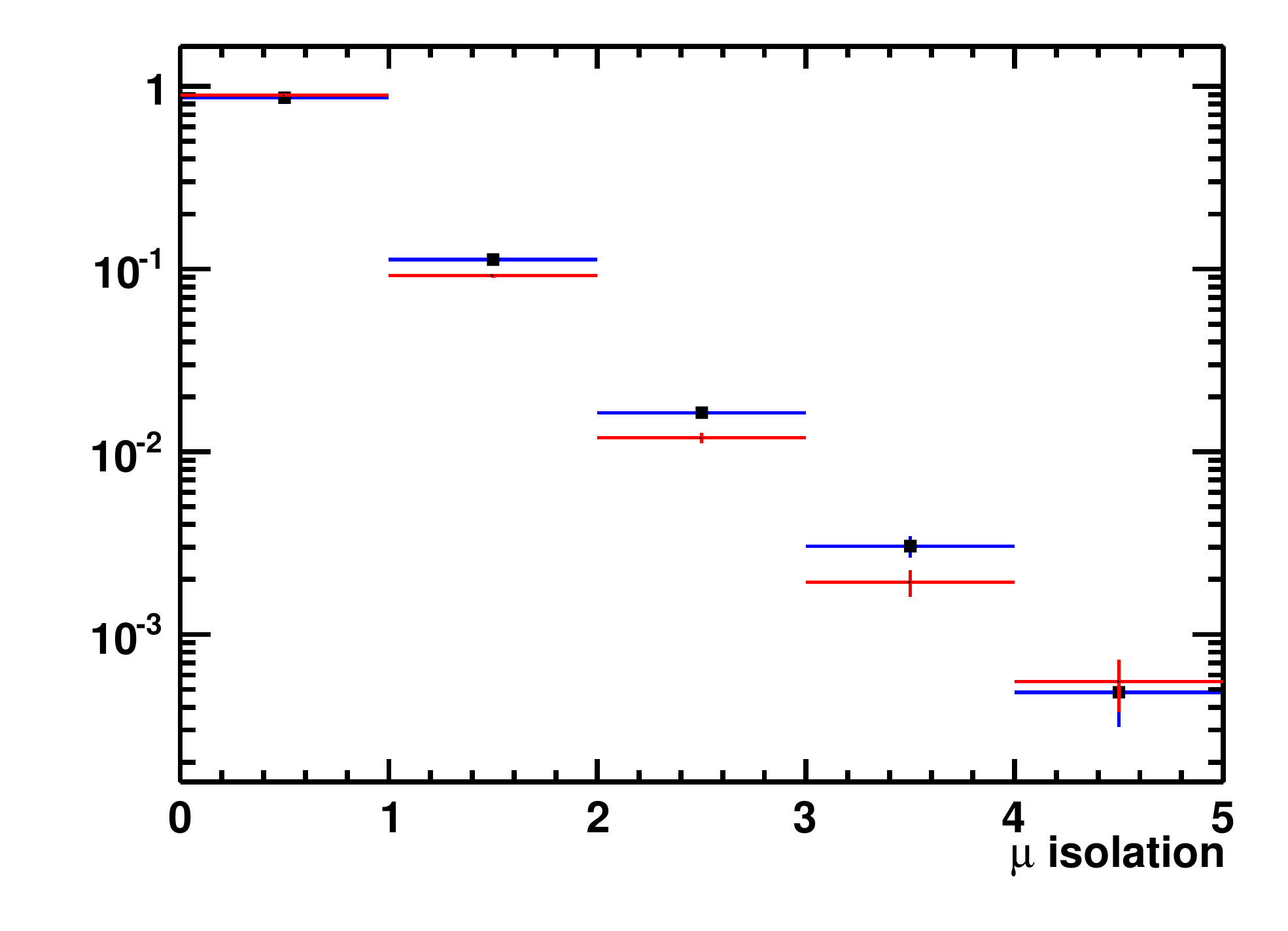}
\caption{Distributions (LHCb) of GL input variables for  $B\to hh'$ (left) and $D^0\rightarrow K^-\pi^+$ (right) decays for data
(blue, black squares) and MC simulation (red). Top plots show the $B$ ($D$) impact parameter, middle plots the DOCA of the two
 tracks, bottom plots the isolation variable of one of the tracks.}
\label{fig:ip_doca_iso}
\end{center}
\end{figure}

Fig.~\ref{fig:gl_vars_1} shows  the input variables to the GL for signal in red and the main background component in black. Such variables, like impact parameters, vertex $\chi^2$, muon isolation etc, are decorrelated before the construction of the geometrical likelihood. Fig.~\ref{fig:ip_doca_iso} shows some of the variables used in the two-body hadronic B decays $B\to hh'$ candidates.
In all cases, some differences are observed
both in the core and in the tails of the distributions. These discrepancies
are more clearly visible in the distributions of the analogous variables for the high statistics
$D^0\rightarrow K^-\pi^+$ control sample, also shown in the same plot.
The optimization and the training of the GL are performed on Monte Carlo (MC) simulated data, but the GL output is then calibrated with data, thanks to the $B\to hh'$ control samples, which have a very similar decay topology, hence the small discrepancies between data and simulation do not introduce a bias into the analysis. The GL is contructed in a way that it should be flat for signal and peaked at 0 for the background.

\begin{figure}[htb]
\centering
\includegraphics[width=0.7\textwidth]{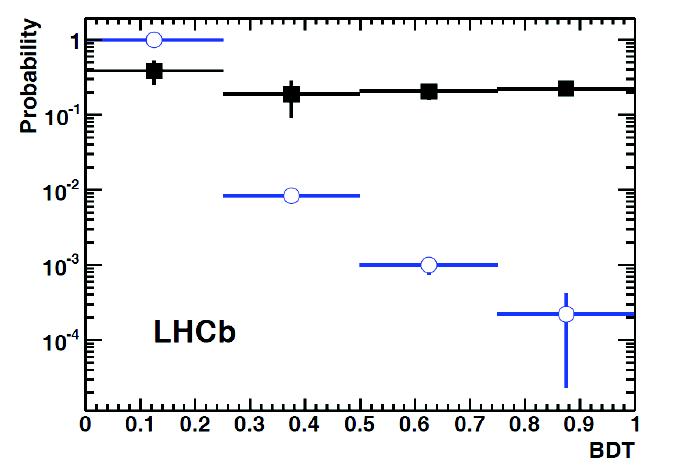}
\caption{GL signal (black full squares) and background (blue empty circles) PDFs used as input for the calculation of the limit (LHCb). }
\label{fig:gl_pdf_bmm}
\end{figure}

Although B->hh' decays represent an excellent proxy for the signal mode, they are in general triggered in a very different maner to the signal, and this biases their kinematical and topological distributions. The solution that has been adopted is to to use only events triggered independently of the signal, that is from other particles in the event, often those from the other b-hadron.
All $B\to hh'$ channels are fit simultaneously, with BR constrained to the PDG values, in bins of GL. The results of the calibration, which also takes care of the small differences between data and simulation, are shown in Fig.~\ref{fig:gl_pdf_bmm} for signal and backfround and it looks very similar to what expected from the simulation within uncertainties. 

The GL PDF for the background is obtained from a fit to the mass sidebands in GL bins. As expected about 97\% of the events are in the first GL bin.

\begin{figure}[htb]
\centering
\includegraphics[width=0.75\textwidth]{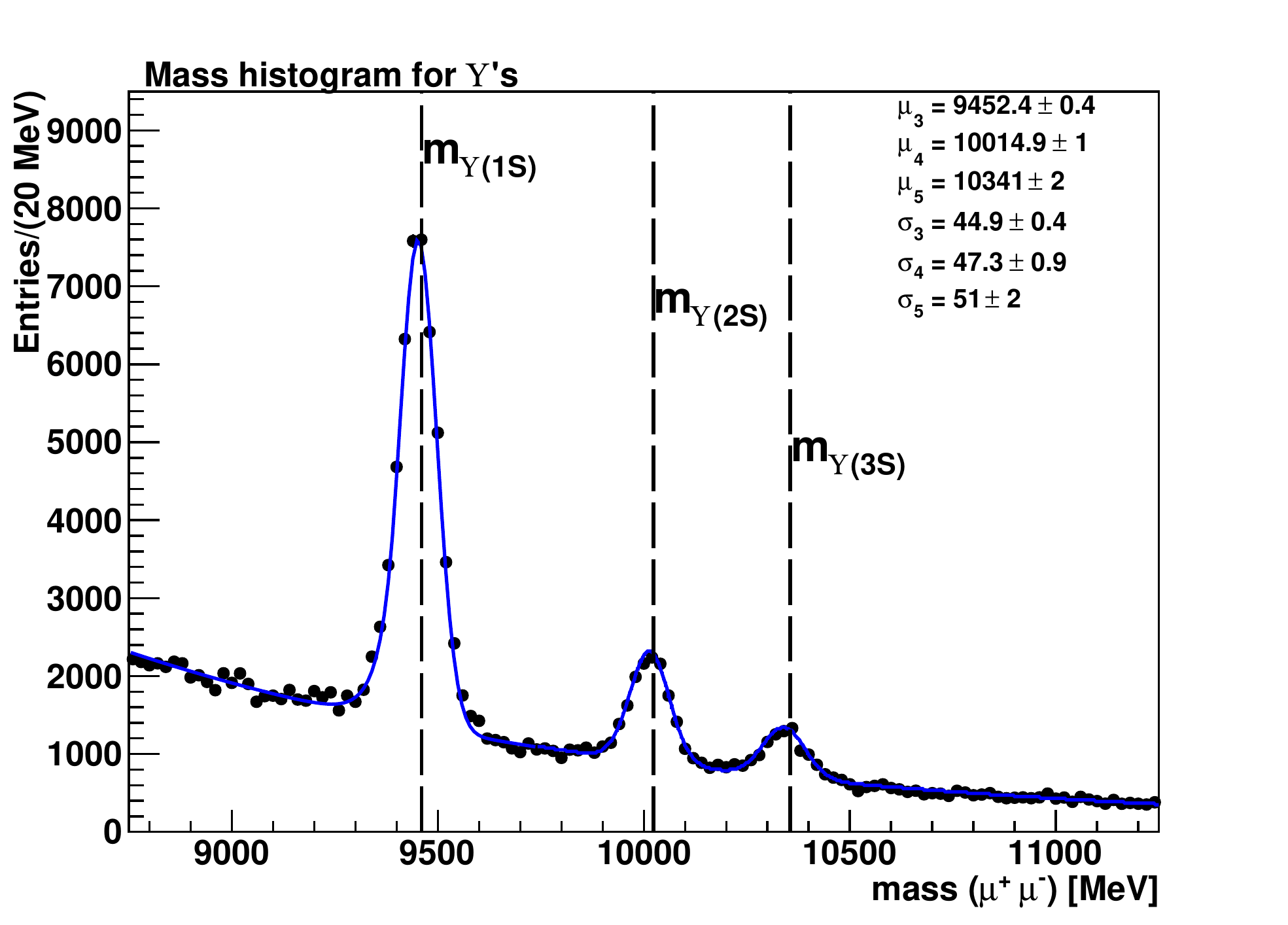}
\caption{
Di-muon invariant mass spectrum between 8.75 GeV/c$^2$ and 11.25 GeV/c$^2$ at LHCb.
The \OneS\ resonance has been fitted with a Crystal Ball function while the \TwoS\ and \ThreeS\
peaks with
Gaussian functions. The background is parametrized with an exponential.}
\label{fig:im2}
\end{figure}

The invariant mass resolution is obtained with two independent methods. The first exploits the fact that, according to the Gluckstern parametrization of relative error of particle's momenta, the resolution is proportional to the mass of the resonances decaying into two muons (after re-weighting the momentum of the muons according to those from $B_s$ mesons). Therefore LHCb extracts the invariant mass resolution at the mass of the $B_s$ from data by linearly interpolating from the measured resolution of charmonium and bottomonium resonances decaying into two muons (i.e. $J/\psi$, $\psi(2S)$\ and \OneS, \TwoS, \ThreeS). Fig.~\ref{fig:im2} shows the $\Upsilon$ family mass distribution, where all resonances are very well separated in LHCb. Another method exploits the  $B\to hh'$ modes inclusively. The two results are compatible and are averaged to give 26.7 MeV/c$^2$. 

For the normalization three different channels have been used: $B^0_{d/s}\to h^+ h'^-$, $B^-\to J/\psi K^-$, and  $B_s\to J/\psi \phi$. Each of these has different efficiencies, different dependencies on the $B_s$ and $B_{d/u}$ hadronization fractions, and different systematic uncertainties. For the ratio of $B^0$ and $B_s$ fragmentation fractons, $f_d/f_s$, which is needed to relate the first two modes to that of the $B_s$ signal decay the HFAG average~\cite{ref:HFAG} of LEP and Tevatron results have been used ($\frac{f_d}{f_s} = 3.71\pm0.47$).

The numbers of found candidates are in agreement with the expected background level, hence LHCb quotes a limit which is 4.3 $\times 10^{-9}$ for 90\% and 5.6 $\times 10^{-9}$  for a 95\% C.L., slightly lower than the expected average as the distribution of the found events are in fact lacking in the very center of the mass distribution, but still compatible with the 68\% expected band.

Similar considerations are valid for the $B^0_d\to \mu^+\mu^-$ mode, where the measured limits are 1.2 and $1.5 \times 10^{-9}$ at the 90 and 95\% C.L.

As for the future, with 200 pb$^{-1}$, a conservative estimate for this summer, LHCb alone could set a 95\% limit down to $2\times 10^{-8}$, while with 1 fb$^{-1}$, expected by the end of the year, it will be able to explore the region of BR down to twice the SM expectation. A 5 standard deviations discovery is expected for BR values down to $10^{-8}$ with the data foreseen in 2012.

The analyses at CMS~\cite{ref:CMS} and ATLAS~\cite{ref:ATLAS} are still blind. They use similar variables as LHCb, except for the fact that they use a B isolation variable rather than a muon isolation one.  The official limit expected for CMS, with the $b\bar b$ cross section currently measured, is $2.3 \times 10^{-8}$ for 1 fb$^{-1}$, but in fact $1.4 \times 10^{-8}$ according to LHCb modified frequentist estimates, hence quite competitive with LHCb. All these expected limits have been calculated assuming only the contribution of events from background.

\section{Asymmetries in $B^0_d\to K^{*0}\mu^+\mu^-$ decays}

Although the decay $B^0_d\to K^{*0}\mu^+\mu^-$ is much more abundant~\cite{ref:kstarmumu_babar} than that of the previous example, it possess high sensitivity to NP through the behaviour of the kinematical observables that can be constructed from the final state distributions, as the Forward Backward Asymmetry ($A_{FB}$) in the $\mu\mu$ rest frame, which will be the focus of the first analyses with this mode in LHC experiments.
What is enticing is that at the zero point of this asymmetry the
dominant theoretical errors due to form factors calculations cancel out.
The analyses are performed blind, using control samples to limit the dependence on simulation in terms of trigger corrections, selection and reconstruction efficiencies, and, especially, acceptance.

\begin{figure}[htb]
\centering
\includegraphics[width=0.7\textwidth]{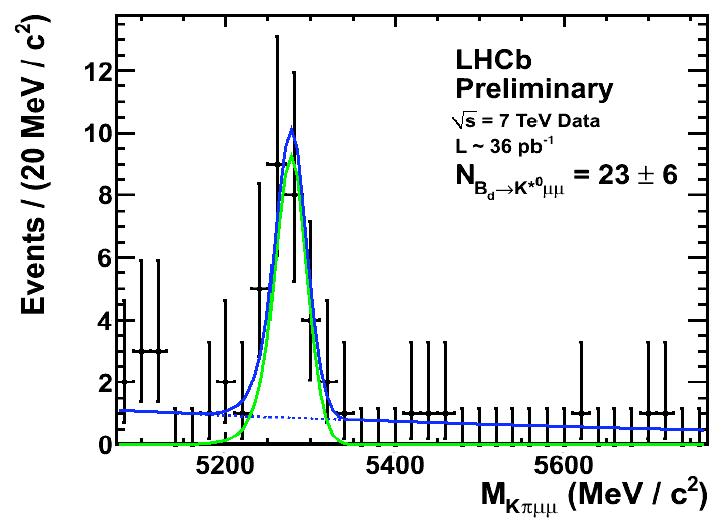}
\caption{LHCb signal for $B^0_d\to K^{*0}\mu^+\mu^-$ candidates}
\label{fig:kstarmumu}
\end{figure}

In LHCb the selection is based on Boosted Decision Trees. The clear signal from 36 pb$^{-1}$ is shown in Fig.~\ref{fig:kstarmumu}. The extrapolated yield for 1 fb$^{-1}$ of data is about 650 candidates with a signal over background ratio even larger than for the B-factories.
All experiments to date see an asymmetry at low $q^2$ opposite, even though statistically not significant, to the SM prediction, a possible hint of NP.  With 1 fb$^{-1}$, assuming LHCb finds the same central value at low $q^2$ as Belle~\cite{ref:BELLE}, by the end of the year we might be faced with an interesting discrepancy of 4 standard deviations with respect to the SM, using only  $B^0_d\to K^{*0}\mu^+\mu^-$ events. 

\section{Conclusions}

LHCb has just published its first $B^0_{s/d}\to\mu^+\mu^-$ search results~\cite{ref:BSMUMU}.  These results are already competitive with the world best limits from Tevatron. LHCb is expected to probe the very interesting BR regions very soon.  New results from Tevatron, but also from CMS and ATLAS, are expected very soon as well.
As far as the $A_{FB}$ for $B^0_d\to K^{*0}\mu^+\mu^-$ decays, LHCb expects to confirm or deny the interesting hints from other experiment by the end of this year.



\end{document}